\documentclass[twocolumn,prb,eqsecnum,showpacs,floatfix,unsortedaddress]{revtex4}
\usepackage{amsmath,mathrsfs}
\usepackage{graphicx}
\begin{document}

\title{Quantum Dissipative Dynamics of the Magnetic Resonance Force
  Microscope in the Single-Spin Detection Limit}

\author{Hanno Gassmann}%
\email{hanno.gassmann@unibas.ch}
\affiliation{Department of Physics and Astronomy, University of Basel,
  Klingelbergstrasse 82, 4056 Basel, Switzerland}%

\author{Mahn-Soo Choi}%
\email{choims@korea.ac.kr} 
\affiliation{Department of Physics, Korea University, Seoul 136-701,
  Korea}%

\author{Hangmo Yi}%
\affiliation{Korea Institute for Advanced Study, 207-43 Cheongryang
  2-dong, Seoul 130-722, Korea}

\author{C. Bruder}%
\affiliation{Department of Physics and Astronomy, University of Basel,
  Klingelbergstrasse 82, 4056 Basel, Switzerland}%

\date{version of September 18th}

\begin{abstract}
We study a model of a magnetic resonance force microscope (MRFM) based
on the cyclic adiabatic inversion technique as a high-resolution tool
to detect single electron spins.  We investigate the quantum dynamics
of spin and cantilever in the presence of coupling to an environment.
To obtain the reduced dynamics of the combined system of spin and
cantilever, we use the Feynman-Vernon influence functional and get
results valid at any temperature as well as at arbitrary system-bath
coupling strength.  We propose that the MRFM can be used as a quantum
measurement device, i.e., not only to detect the modulus of the spin
but also its direction.
\end{abstract}

%03.65.Yz Decoherence; open systems; quantum statistical methods (see also
%         03.67.Pp in quantum information; for decoherence in Bose-Einstein
%         condensates, see 03.75.Gg)
%07.79.Pk Magnetic force microscopes
%07.10.Cm Micromechanical devices and systems (for micro- and
%         nano-electromechanical systems (MEMS/NEMS), see 85.85.+j in
%         electronic and magnetic devices)
\maketitle

%%%%
\let\eps=\epsilon%
\let\up=\uparrow%
\let\down=\downarrow%
\newcommand\eff{\mathrm{eff}}%
\newcommand\tot{\mathrm{tot}}%
\newcommand\bfB{\mathbf{B}}%
\newcommand\varA{\mathscr{A}}%
\newcommand\varD{\mathscr{D}}%
\newcommand\varH{\mathscr{H}}%
\newcommand\varO{\mathscr{O}}%
\newcommand\varU{\mathscr{U}}%
\newcommand\varW{\mathscr{W}}%
\newcommand\varZ{\mathscr{Z}}%
\newcommand\hatb{\hat{b}}%
\newcommand\half{\frac{1}{2}}%
\newcommand\ket[1]{\left|\textstyle{#1}\right\rangle}%
\newcommand\bra[1]{\left\langle\textstyle{#1}\right|}%
\newcommand\avg[1]{\left\langle\textstyle{#1}\right\rangle}%
\newcommand\im{{\,\mathrm{Im}\,}}%
\newcommand\re{{\,\mathrm{Re}\,}}%
\newcommand\tr{\mathrm{tr}}%

\section{Introduction}
Magnetic resonance imaging technologies (MRI, NMR, ESR) are widely
used to characterize physical, chemical, and biological samples.  What
makes them powerful is that they are non-destructive and capable to
probe the three-dimensional structure of the
sample~\cite{Slichter90a}.  Recently, looking at structures at the
molecular or atomic level has become important in a number of
scientific disciplines. Magnetic resonance force microscopes (MRFMs)
have been developed to bring magnetic resonance imaging technologies
to such an ultimate resolution.  The MRFM combines conventional
magnetic resonance technology with probe microscope technology, e.g.,
atomic force microscopy, to image individual molecules or
atoms~\cite{Sidles95a}.  In an MRFM, a magnetic particle mounted on a
cantilever interacts with nuclear or electron spins in the sample via
the very weak magnetic dipole force.  When modulated at resonance with
the cantilever oscillation frequency, even the weak magnetic force
induces sufficiently large vibrations of the cantilever.  By probing
the resulting vibrational motion of the cantilever, it is in principle
possible to detect spins with molecular or atomic resolution.  The
cyclic adiabatic inversion (CAI) technique has been
proposed~\cite{Sidles95a} as a promising method to modulate the
magnetic force.

The future of the MRFM depends crucially on the development of proper
mechanical micro-resonators, e.g., cantilevers~\cite{Bocko96a}.
Remarkable progress has been made in this direction and the detection
of atto-newton or subatto-newton scale forces has been achieved
already\cite{Mamin01a,Stowe97a}. Recently, a nanomechanical flexural
resonator at microwave frequencies has also been
realized~\cite{Huang03a}.
The development of the proper technology to detect nanometer-scale
mechanical motion is also important.  Optical interferometry or
electrical parametric transducers are the most common
examples~\cite{Mamin01a,Stowe97a,Abramovici96a}.  In recent work, a
single-electron transistor capacitively coupled to a nanomechanical
resonator has been used to detect the vibrational motion of the
resonator even in the quantum regime~\cite{Knobel03a}.

The progress in MRFM and related technologies has also attracted
theoretical interest, especially, the question of single-spin 
detection using the MRFM.  
Mozyrsky \textit{et al.}~\cite{Mozyrsky03} studied the
relaxation of a spin, treating the cantilever as a classical noise source.
Berman \textit{et al.}~\cite{Berman03a,Berman03b} studied a CAI-based
MRFM and treated both the spin and the cantilever as quantum systems
that are subject to environmental effects.  They addressed two
interesting and important issues: which component is measured in an
MRFM single-spin measurement and whether the two spin states (up and
down) lead to distinctively different cantilever motions.  They solved
numerically the time-dependent Schr\"odinger equation for the
spin-plus-cantilever system in the absence of coupling to the
environment.  In the presence of an environment, they constructed a
generalized master equation in the high-temperature limit, and solved
it numerically.  We note that their master equation is based on the
Markov approximation, and is not in Lindblad
form~\cite{Lindblad76a,Diosi93a} (the normalization and the positivity
of the density matrix are not guaranteed).

In this paper, we study the measurement of single spins with the MRFM
based on the CAI technique.  The starting point of our work is closely 
related to Refs.~\onlinecite{Berman03a,Berman03b}. 
In the absence of the coupling to the environment, we solve
the time-dependent Schr\"odinger equation exactly and confirm the
numerical results by Berman \textit{et al.}~\cite{Berman03a,Berman03b} 
We use an open quantum system 
approach~\cite{Breuer02a,Weiss00a}, i.e., we take the influence of
the environment into account by coupling a harmonic oscillator bath to
the cantilever.
To calculate the dynamics of the spin
during the measurement process, we take an effective-bath approach,
and obtain the exact solution for the reduced density matrix of the
spin.
To find the cantilever dynamics, we solve the Feynman-Vernon influence
functional~\cite{feynmanvernon,grabert} 
in order to obtain the reduced density matrix of the spin
plus cantilever system.
Both methods are are valid at any temperature as
well as for arbitrary coupling strength (within the CAI-scheme).
This analytical approach allows
us to interpret the results in a transparent way and to investigate the
issue whether the MRFM can be used as a quantum measurement device
to probe the spin state.

The paper is organized as follows: In Section~\ref{mrfm::sec:model} we
first introduce the model and discuss our adiabatic Born-Oppenheimer
approximation scheme in connection with the CAI-technique. In
Section~\ref{mrfm::sec:noenv} we present the exact solution of the
time-dependent Schr\"odinger equation for the spin-plus-cantilever
system without coupling to the environment, the results of which will be
compared with those in the dissipative case in the later sections.  In
Section~\ref{mrfm::sec:spin}, we investigate the quantum dissipative
dynamics of the spin alone using an effective-bath approach.  The
dynamics of the cantilever is investigated in
Section~\ref{mrfm::sec:cantilever}.  The physical
implications of the solution are analyzed in detail and the possibility
to use the MRFM as a quantum measurement device
is discussed in Section~\ref{mrfm::sec:qmeasure}.  Finally,
in Section~\ref{mrfm::sec:conclusion} we draw our conclusions.

\begin{figure}
\centering%
\includegraphics*[width=0.65\linewidth]{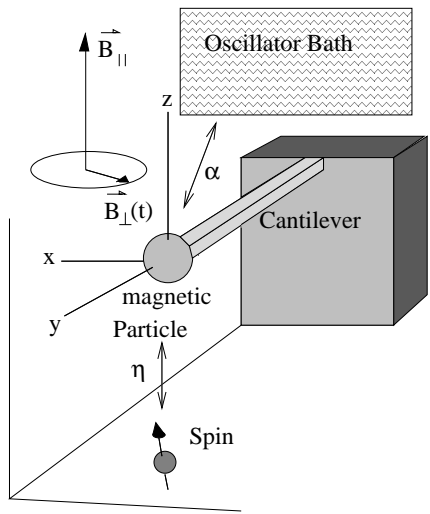}
\caption{\label{fig:setup} MRFM measurement device.
A cantilever carrying a magnetic particle is subject to a static
magnetic field $\mathbf{B}_\|$ in the $z$-direction, and a
time-dependent field $\mathbf{B}_\bot(t)$ rotating with
frequency $\omega_\mathrm{rf}$ in the $x$-$y$ plane. The cantilever is
coupled to a sample spin by a magnetic force $\eta$.}
\end{figure}

\section{\label{mrfm::sec:model}Model}

We consider an MRFM setup based on the cyclic adiabatic inversion (CAI)
technique (see Fig.~\ref{fig:setup}).  It consists of a ferromagnetic
particle mounted on the tip of a cantilever, a strong static magnetic
field $\mathbf{B}_\|$ in the $z$-direction, and an 
rf field $\mathbf{B}_\bot(t)$ rotating with
frequency $\omega_\mathrm{rf}$ in the $x$-$y$ plane modulated by $\phi(t)$:
\begin{equation}
\mathbf{B}_\bot(t)=B_\bot\left[
\begin{array}{c}
\cos\left(\omega_\mathrm{rf}t-\phi(t)\right)\\
-\sin\left(\omega_\mathrm{rf}t-\phi(t)\right)
\end{array}
\right] \,.
\end{equation}
As in usual NMR setups, one puts
\begin{math}
\omega_\mathrm{rf} = \eps_z \equiv g\mu_BB_\|
\end{math},
where $g$ is the $g$-factor of the spin and $\mu_B$ is the Bohr magneton.
For later use, we also define
\begin{math}
\eps_\bot \equiv g\mu_BB_\bot
\end{math}.
The ``sample'' consists of a spin interacting with the ferromagnetic
particle via the magnetic force $\eta$ and with the static and rf
fields.  The Hamiltonian of the spin and the cantilever is given by
\begin{multline}
\label{mrfm::eq:H:0}
\varH(t) = -\frac{\eps_z}{2}\hat\sigma_z
- \frac{\eps_\bot}{2}\left[
  \hat\sigma_+ e^{i\eps_z{t}-i\phi(t)}
  + h.c.
\right] \\
- \eta\,\hat\sigma_z\hat{z}
+ \frac{\hat{p}_z^2}{2}
+ \frac{\hat{z}^2}{2} \,,
\end{multline}
where the $\hat\sigma$'s are Pauli matrices,
\begin{math}
\hat\sigma_\pm=(\hat\sigma_x\pm i\hat\sigma_y)/2
\end{math}, and $\hat{z}$ ($\hat{p}_z$) is the position (momentum) operator
of the cantilever. 
In Eq.~\eqref{mrfm::eq:H:0} and hereafter we use a unit system
such that $\hbar=k_B=\omega_0=\ell_0=1$, where $\omega_0$ is the natural
frequency of the cantilever and $\ell_0\equiv\sqrt{\hbar/m\omega_0}$ is
the harmonic-oscillator length.  It is convenient to move to a frame
rotating with the rf field by making a transformation\cite{Messiah61b}
\begin{equation}
\label{mrfm::eq:A}
\varH \to \varA^\dag\varH\varA - i\varA^\dag\dot\varA
\end{equation}
with
\begin{math}
\varA = \exp\left\{\frac{i}{2}[\eps_zt-\phi(t)]\hat\sigma_z\right\}
\end{math}.
The resulting Hamiltonian reads\cite{Berman03a,Berman03b}
\begin{equation}
\label{mrfm::eq:H:1}
\varH(t) = 
- \frac{1}{2}\dot{\phi}(t)\hat{\sigma}_{z}
- \frac{1}{2}\eps_\bot\hat{\sigma}_{x}
- \eta\,\hat{\sigma}_{z}\hat{z}
+ \frac{\hat{p_{z}}^2}{2}+\frac{\hat{z}^2}{2} \,.
\end{equation}
The idea of the CAI-based MRFM is as follows: The phase modulation
$\phi(t)$ of the rf field is assumed to be harmonic and causes
adiabatic inversions of the spin, which in turn exert an oscillating
force on the cantilever.
At resonance, i.e., if the frequency of the modulation is equal to
the natural frequency of the cantilever (which is $1$ in our units),
\begin{equation}
\label{mrfm::eq:dotphi}
\dot\phi(t) = \phi_0\sin(t - \varphi) \,,
\end{equation}
the vibration amplitude of the cantilever can be large even for a
very small magnetic force $\eta$.

Equation~\eqref{mrfm::eq:H:1} describes a spin which couples to a harmonic
oscillator and is itself subject to a time-dependent effective magnetic
field
\begin{math}
g\mu_B\bfB_\eff(t)
\equiv \eps_\bot\mathbf{e}_x + \dot\phi(t)\mathbf{e}_z
\end{math}, 
where $\mathbf{e}_x$, $\mathbf{e}_z$ are unit vectors in the rotating
system. 
The Hamiltonian in Eq.~\eqref{mrfm::eq:H:1} is not exactly
solvable.  Here we make a plausible approximation based on the following
observations.  For typical experimental
parameters\cite{Berman03a,Berman03b}, 
$\bfB_\eff$ varies slowly compared with the Rabi
oscillation frequency:
\begin{math}
|\dot{\bfB}_\eff(t)|/|\bfB_\eff(t)|
\ll \eps(t)
\equiv \sqrt{\eps_\bot^2 + \dot\phi^2(t)}
\end{math}.
According to the adiabatic theorem~\cite{Avron87a,Messiah61b}, the spin
part of the solution should be determined by the adiabatic evolution; i.e.,
the spin ``follows adiabatically'' the effective field $\bfB_\eff(t)$.
It is therefore convenient to choose the basis states $\ket{\chi_+(t)}$
and $\ket{\chi_-(t)}$ quantized along the axis parallel to
$\bfB_\eff(t)$ (notice that there is no Berry phase because the solid
angle enclosed by $\bfB_\eff(t)$ is zero).  In this basis, the
Hamiltonian in Eq.~\eqref{mrfm::eq:H:1} is recast to
\begin{multline}
\label{mrfm::eq:H:2}
\varH(t)
= -\frac{1}{2}\epsilon(t)\hat{\tau}_{z}
- \eta\frac{\dot\phi(t)}{\eps(t)}\hat{\tau}_{z}\hat{z}
+ \eta\frac{\eps_\bot}{\eps(t)}\hat{\tau}_{x}\hat{z} \\\mbox{}
+ \frac{\hat{p}_{z}^2}{2}+\frac{\hat{z}^2}{2} \,,
\end{multline}
where \(\hat{\tau}_{x}\) and \(\hat{\tau}_{z}\) are the Pauli matrices with
respect to the frame rotating adiabatically with 
$\bfB_\eff(t)$.
We have suppressed the time arguments to the Pauli matrices
\begin{math} \hat\tau_z(t) \equiv \ket{\chi_+(t)}\bra{\chi_+(t)} -
\ket{\chi_-(t)}\bra{\chi_-(t)} \end{math} and \begin{math}
\hat\tau_x(t) \equiv \ket{\chi_+(t)}\bra{\chi_-(t)} +
\ket{\chi_-(t)}\bra{\chi_+(t)} \end{math} in
Eq.~\eqref{mrfm::eq:H:2}. This is because within the adiabatic
approximation, the dynamics of the spin part of the wave function
is completely governed by the basis states $\ket{\chi_\pm(t)}$ and
the dynamic phases, i.e., 
\begin{math} \ket{\chi(t)} =
c_+e^{-i\int_0^{t}{dt'}\eps_+(t')}\ket{\chi_+(t)} +
c_-e^{-i\int_0^{t}{dt'}\eps_-(t')}\ket{\chi_-(t)} 
\end{math}. 
We further note that the spin dynamics is much faster than the
cantilever motion,
\begin{math}
\eps(t) \geq \eps_\bot \gg 1
\end{math}.
The situation is reminiscent of the Born-Oppenheimer
approximation\cite{Ashcroft76a}, where the nuclei interact with the
average charge density of the electrons which move much faster. In our
system the nuclei correspond to the harmonic oscillator which is
interacting with the averaged motion of the spin.  Therefore one can
drop the third term in Eq.~\eqref{mrfm::eq:H:2}. (The deviation of the
spin due to this term is also negligibly small since
$\eta|\avg{\hat{z}(t)}|\ll\eps(t)$; see below).  Using this approximation we
finally get the following Hamiltonian, which is the basis of the
further considerations in the paper:
\begin{equation}
\label{mrfm::eq:H:3}
\varH(t)
= -\frac{1}{2}\eps(t)\hat{\tau}_{z}
- \eta f(t)\hat{\tau}_{z}\hat{z}
+ \frac{\hat{p}_{z}^2}{2}+\frac{\hat{z}^2}{2} \,,
\end{equation}
where
\begin{math}
f(t) \equiv \dot\phi(t)/\eps(t)
\end{math}.
This form is justified in a more rigorous way in
Appendix~\ref{mrfm::sec:Landau-Zener}, also taking into account the
influence of the environment (see below).  Its validity was also
confirmed by the numerical simulations in Ref.~\onlinecite{Berman03b}.

So far we have described a model for an idealized system of spin and
cantilever.  In reality they are coupled to various environments,
which lead to decoherence as well as damping.  In particular, the
cantilever is inevitably under the influence of phonons or other
vibrational modes which are close in frequency to the single mode in
question.  The (direct) environmental effects for the spin, e.g.,
hyperfine interaction, spin-lattice relaxation, etc., are relatively
small.  Therefore, for simplicity, we assume a simple Ohmic bath of
oscillators~\cite{Caldeira83b,Caldeira83a,Caldeira81a,Weiss00a} directly
coupled to the cantilever but not to the spin.  Then the total
Hamiltonian for the spin and the cantilever plus the oscillator bath is
given by
\begin{multline}
\label{mrfm::eq:H:total}
\varH_\mathrm{total}(t)
= \varH(t) \\\mbox{}
+ \sum_{k=1}^{\infty}\left[\frac{\hat{p}_{k}^2}{2 m_{k}}
  + \frac{m_{k}\omega_{k}^2}{2}
  \Big(\hat{x}_{k}-\frac{c_{k}}{m_{k}\omega_{k}^2}\hat{z}\Big)^2\right]\,.
\end{multline}
All the relevant features of the Ohmic bath are characterized by the
spectral density
\begin{eqnarray}
\label{mrfm::eq:J}
J(\omega) &=& \frac{\pi}{2} \sum_k \frac{c_k^2}{m_k\omega_k}
\delta(\omega-\omega_k) \nonumber\\ 
&=& \alpha\omega\Theta(1-\omega/\omega_C) \,,
\end{eqnarray}
where $\alpha$ is a dimensionless parameter characterizing the
coupling between the system and the environment and $\omega_C$ is the
cut-off frequency. The spin dynamics and the probability
distribution of the cantilever will not depend on the cut-off.

We describe the system of spin plus cantilever in terms of the reduced
density matrix $\hat\rho(t)\equiv\tr_B\hat\rho_\tot(t)$ by tracing out
the bath.  In the realistic typical experimental situation, the
cantilever always remains in contact with the environment. Thus, the
cantilever and bath are not in a product state at the beginning of the
experiment. For the calculation with the influence functional, we can
take this fact into account, assuming that the cantilever and the bath
were in a factorized state at a time \(t=t_{0}\). In the limit
\(t_{0}\rightarrow -\infty\) we get then the realistic initial state
for the cantilever at the time \(t=0\). If we would start with a
factorized state between cantilever and bath, the solution would be
very sensitive to the initial condition of the cantilever; see
Section~\ref{mrfm::sec:spin}.

Furthermore, it is assumed that the interaction between the spin and the
cantilever is turned on at $t=0$, i.e., \(f(t)=0\) for \(t<0\). The
measurement happens at times \(t>0\). The initial state $\hat\rho(0)$ of
the density matrix is a product state,
\begin{equation}
\label{mrfm::eq:W(0)}
\hat\rho(0) = \hat\rho^{(S)}(0)\hat\rho^{(C)}(0) \,,
\end{equation}
where $\hat\rho^{(S)}$ is the density matrix for
the spin only and $\hat\rho^{(C)}$ describes the cantilever in thermal
equilibrium with the bath. From the CAI scheme and from the
associated adiabatic approximation discussed above it then follows
that the density matrix at times $t>0$ has the form
\begin{equation}
\label{mrfm::eq:W(t)}
\bra{s,z}\hat\rho(t)\ket{s',z'} =
\rho_{ss'}^{(S)}(0)\rho_{ss'}^{(C)}(z,z',t) \,.
\end{equation}
Namely, the dynamics of the density matrix $\hat\rho(t)$ is completely
determined by the spin-dependent cantilever part
$\rho_{ss'}^{(C)}(z,z',t)$.

Here the spin-dependent cantilever part should not be confused with the
density matrix for the cantilever only, which is given by
\begin{multline}
\rho^{(C)}(z,z',t)
= \sum_{s=\pm}\bra{s,z}\hat\rho(t)\ket{s,z'} \\
= \rho_{++}^{(S)}(0)\rho_{++}^{(C)}(z,z',t)
+ \rho_{--}^{(S)}(0)\rho_{--}^{(C)}(z,z',t) \,.
\end{multline}
Analogously, the density matrix for the spin only at time $t>0$ is given
by
\begin{equation}
\rho_{ss'}^{(S)}(t)
= \rho_{ss'}^{(S)}(0)\int_{-\infty}^{\infty}{dz}\;\rho_{ss'}^{(C)}(z,z,t) \,.
\end{equation}

There are several ways to prepare the spin in a particular
state~\cite{Weiss00a}, and we will assume a general state
$\rho_{ss'}^{(S)}(0)$.

\section{\label{mrfm::sec:noenv}
  The coherent solution without bath}

Before we investigate the full Hamiltonian in
Eq.~\eqref{mrfm::eq:H:total}, it will be instructive to first consider
the problem without bath, Eq.~\eqref{mrfm::eq:H:3}.  The time-dependent
Hamiltonian in Eq.~\eqref{mrfm::eq:H:3} can be solved exactly for
arbitrary functions $\eps(t)$ and $f(t)$ of $t$ (of course, the
variation of $\eps(t)$ and $f(t)$ in time should be sufficiently slow
so that the Hamiltonian Eq.~\eqref{mrfm::eq:H:3} is meaningful).

One can show that the time-evolution operator
\begin{math}
\varU(t_{2},t_{1})
\equiv \widehat{T}\exp\left[-i\int_{t_1}^{t_{2}}dt' \varH(t')\right]
\end{math}
($\widehat{T}$ is the time-ordering operator) is given by
\begin{multline}
\label{mrfm::eq:U(t)}
\varU(t_2,t_1)
= \exp\left[ic(t_1,t_2)+
\frac{i}{2}\int_{t_1}^{t_2}{dt'}\;\eps(t')\, \hat\tau_z
\right] \\\mbox{}\times
\varD(\hat\tau_z\xi(t_2))\varU_0(t_2-t_1)
\varD^\dag(\hat\tau_z\xi(t_1)) \,,
\end{multline}
where
\begin{equation}
\label{mrfm::eq:xi}
\xi(t)
\equiv i\eta\frac{1}{\sqrt{2}}\int_{0}^{t}{dt'}\; e^{-i(t-t')}f(t') \,,
\end{equation}
\begin{equation}
\label{mrfm::eq:U0}
\varU_{0}(t) \equiv \exp\left( -i t\hat{a}^{\dagger}\hat{a} \right) \,,
\end{equation}
$\hat{a}=(\hat{z}+i\hat{p}_z)/\sqrt{2}$,
and $\varD(\xi)$ is a displacement operator\cite{Gardiner00a} defined
for a complex number $\xi$ by
\begin{equation}
\label{mrfm::eq:D}
\varD(\xi) = \exp(\xi\hat{a}^\dag - \xi^*\hat{a}) \,.
\end{equation}
The coefficient \(c(t_{1},t_{2})\) in Eq.~\eqref{mrfm::eq:U(t)} is a real
function of $t_1$ and $t_2$ (one does not need an explicit expression of
it because it drops out of the following calculations).

To illustrate the dynamics created by the time-evolution operator in
Eq.~\eqref{mrfm::eq:U(t)}, let us discuss an example.
Suppose that we start at time
\(t=0\) with the cantilever in a coherent state
\begin{equation}
\psi(z,0)
= \frac{1}{\sqrt[4]{\pi}}
\exp\left[
  -\half z^2 + \sqrt{2}\xi_0 z
  -(\re\xi_0)^2
\right]
\end{equation}
and with the spin in a linear superposition (with amplitudes $c_+$ and $c_-$)
\begin{equation}
\label{mrfm::eq:Psi(0)}
\ket{\chi(0)} = c_{+}|\chi_{+}(0)\rangle
  + c_{-}|\chi_{-}(0)\rangle\,.
\end{equation}
The total wave function at $t=0$ is given by
\begin{equation}
\ket{\Psi(z,0)}
=\psi(z,0)\ket{\chi(0)} \,,
\end{equation}
and at later time $t>0$, by
\begin{equation}
\label{mrfm::eq:Psi(t)}
\ket{\Psi(z, t)} = 
c_+\psi_{+}(z,t)|\chi_{+}(t)\rangle
+ c_-\psi_{-}(z,t)|\chi_{-}(t)\rangle \,.
\end{equation}
The cantilever wave function in Eq.~\eqref{mrfm::eq:Psi(t)} for each spin
component is given by
\begin{multline}
\psi_\pm(z,t) = \frac{1}{\sqrt[4]{\pi}}
\exp\left[ic(t,0)\pm i\int_0^t{dt'}\;\eps(t')\right] \\\mbox{}\times
\exp\left\{-\half z^2 + \sqrt{2}\xi_\pm'(t)z
  - [\re\xi_\pm'(t)]^2\right\} \,,
\end{multline}
where
\begin{equation}
\label{mrfm::eq:xi'}
\xi_\pm'(t)
= \pm\xi(t) + \xi_0 e^{-it} \,.
\end{equation}
Therefore, the average position of the cantilever is
\begin{math}
\avg{\hat{z}(t)}_\pm = \sqrt{2}\re\xi_\pm'(t)
\end{math}
for spin $s=\pm$, respectively, whereas the average momentum is given by
\begin{math}
\avg{\hat{p}_z(t)}_\pm = \sqrt{2}\im\xi_\pm'(t)
\end{math}.
Here it is interesting to note (in comparison with the results below)
that exactly at resonance [see Eq.~\eqref{mrfm::eq:dotphi}],
$|\xi(t)|$ in Eq.~\eqref{mrfm::eq:xi} [and hence $|\xi_\pm'(t)|$ in
Eq.~\eqref{mrfm::eq:xi'}] contains a term which linearly increases
with time $t$.  In other words, the oscillation amplitude of the
cantilever gets indefinitely larger and larger as time passes.  This
is not surprising since we are driving an \emph{ideal} oscillator
at the resonance frequency, and in fact this is what allows the MRFM
to detect ultra-small forces.  In reality, the cantilever is subject
to various environmental effects and the oscillation amplitude is
bounded from above (i.e., the Q-factor is finite). This is the case
that we will study below.

\section{\label{mrfm::sec:spin}Dynamics of the Spin}

Now we take into account the influence of the bath.  In this section, we
first analyze the dynamics of the spin.  The dynamics of the cantilever
will be discussed in the following section.  When we are interested in
the dynamics of the spin alone, we can regard the cantilever as a part
of the environment.
In fact, Garg \textit{et al.}~\cite{Garg85a} (see also
Refs.~\onlinecite{Gassmann02,Wilhelm03,kleff}) showed that the problem
is equivalent to a spin coupled linearly to an oscillator bath:
\begin{multline}
\label{mrfm::eq:H:total:2}
\varH_\tot(t)
= -\frac{1}{2}\epsilon(t)\hat{\tau}_{z}
- \eta f(t)\hat{\tau}_{z}\sum_kg_k\left(\hatb_k^\dag + \hatb_k\right)
\\\mbox{}
+ \sum_k\omega_k\hatb_k^\dag\hatb_k \,.
\end{multline}
The distribution of the oscillator frequencies $\omega_k$ and the
coupling constant $g_k$ are now characterized by
a non-Ohmic spectral density
\begin{equation}
J_\eff(\omega)
\equiv \sum_kg_k^2\delta(\omega-\omega_k)
= \frac{1}{\pi}\,
\frac{\alpha\omega}{(\omega^2-1)^2+(\alpha\omega)^2}\,.
\end{equation}

To investigate the spin dynamics, we write the reduced density matrix of
the spin
\begin{equation}
\hat\rho^{(S)}(t)
= \tr_B\varU_\tot(t)\hat\rho_\tot(0)\varU_\tot^\dag(t)
\end{equation}
in terms of the time-evolution operator $\varU_\tot(t)$ associated with
$\varH_\tot(t)$ in Eq.~\eqref{mrfm::eq:H:total:2}. In analogy to
Eq.~\eqref{mrfm::eq:U(t)}, the time-evolution operator is given by
\begin{equation}
\varU_\tot(t)
= \exp\left[\frac{i}{2}\int_0^t{dt'}\eps(t')\hat\tau_z\right]
\prod_k \varD(\hat\tau_z\xi_k(t))
e^{-i\omega_k t\hatb_k^\dag \hatb_k} \,,
\end{equation}
where
\begin{equation}
\xi_k(t)=i\eta g_k \int_0^t dt' e^{-i(t-t')\omega_k}f(t')\,,
\end{equation}
and \(\varD\) is now the displacement operator for the $k$-th mode of
the bath, i.e., $\hat{a}$ should be replaced by $\hat{b}_k$ in 
Eq.~(\ref{mrfm::eq:D}).

For the initial state $\hat\rho_\tot(0)$, we assume [see
Eqs.~\eqref{mrfm::eq:W(0)}]
\begin{equation}
\hat\rho_\tot(0)
= \hat\rho^{(S)}(0)
\prod_k
\frac{e^{-\beta\omega_k\hatb_k^\dag\hatb_k}}{Z_k} \,.
\end{equation}
Then the density matrix for the spin is given by
\begin{multline}
\label{mrfm::eq:rho(t):spin}
\rho_{ss'}^{(S)}(t)
= \rho_{ss'}^{(S)}(0)
\exp\left[i\frac{(s-s')}{2}\int_0^t{dt'}\eps(t')\right]
\\\mbox{}\times
\prod_k\avg{
  \varD^\dag(s'\xi_k(t))\varD(s\xi_k(t))
}_k \,,
\end{multline}
where $\avg{\cdots}_k$ is the average with respect to the $k$-th
oscillator in the bath.

Equation~\eqref{mrfm::eq:rho(t):spin} shows that the diagonal elements
of the density matrix ($s=s'$) are constant in time
\begin{equation}
\rho_{ss}^{(S)}(t)
= \rho_{ss}^{(S)}(0) \,.
\end{equation}
In other words, there is no spin relaxation and the spin dynamics is
pure dephasing because there are no transverse fields.  This is
consistent with the adiabatic approximation we made at the beginning.

\begin{figure}
\centering%
\includegraphics*[width=75mm]{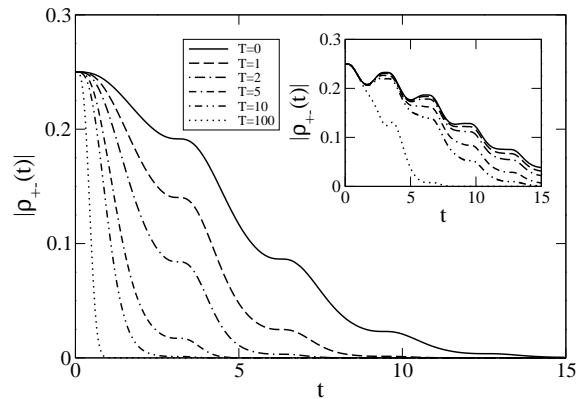}
\caption{\label{mrfm::fig:2} Main plot: \(|\rho_{+ -}(t)|\) for
  different temperatures \( T=0,1,2,5,10,100\),
  for \(\phi_0=1000\), \(\eps_\bot=400\), \(\eta=0.3\),
  \(\alpha=0.006\), and $\omega_C=1000$.  The initial condition for
  cantilever and bath is the thermal equilibrium state.
  Inset: same quantity for an initial product state of cantilever and
  bath. Initially, the cantilever wave function is a Gaussian with width
  \(\sigma=\sqrt{2}\).  In both cases,
  $\rho_{ss'}^{(S)}(0)=1/2$ for $s,s'=\pm$.}
\end{figure}

On the other hand, the off-diagonal elements ($s\neq s'$) are expected
to vanish rapidly with time.  This can be seen from [see
Eq.~\eqref{mrfm::eq:rho(t):spin}]
\begin{equation}
\label{mrfm::eq:rho(t):spin:2}
\rho_{+-}^{(S)}(t)
= \rho_{+-}^{(S)}(0)\exp\left[-\Gamma(t)
  + i\int_0^t{dt'}\eps(t')\right] \,,
\end{equation}
where
\begin{equation}
\Gamma(t) \equiv
2\sum_k|\xi_k(t)|^2 \coth\Big(\frac{\omega_k}{2 T}\Big) \,,
\end{equation}
or in terms of the spectral density function
\begin{equation}
\label{mrfm::eq:Gamma(t):2}
\Gamma(t)
= 2\eta^2\int_0^\infty{d\omega}J_\eff(\omega)\coth\Big(\frac{\omega}{2
T}\Big)
\left|\int_0^t{dt'}e^{i\omega t'} f(t')
\right|^2 \,.
\end{equation}
Figure~\ref{mrfm::fig:2} shows $|\rho_{+-}(t)|$ evaluated using
Eqs.~\eqref{mrfm::eq:rho(t):spin:2} and \eqref{mrfm::eq:Gamma(t):2}.
To compare our results with those of Berman \textit{et
al.}~\cite{Berman03a,Berman03b} who assumed an initial product state
of cantilever and bath, the inset of Fig.~\ref{mrfm::fig:2} shows
$|\rho_{+-}^{(S)}(t)|$ for a Gaussian initial state of the cantilever.
(To obtain these results we evaluate the path-integral formulas in
Appendix~\ref{mrfm::sec:path} with \(t_{0}=0\) instead of taking the
limit \(t_{0}\rightarrow -\infty\) ).  If we compare the main part of
Fig.~\ref{mrfm::fig:2} with the inset the strong dependence on the
initial conditions is evident.  The slower decay and the more
pronounced oscillations shown in the inset are a consequence of the
oscillatory relaxation of the cantilever to its thermal equilibrium
state if one starts with an initial product state of cantilever and
bath. On increasing the coupling \(\alpha\), the oscillatory behavior
becomes less visible since the cantilever relaxes immediately to its
thermal state.

\section{\label{mrfm::sec:cantilever}Dynamics of the Cantilever}

In Section~\ref{mrfm::sec:noenv}, we described the driven dynamics of
the otherwise isolated system of spin and cantilever determined by the
Hamiltonian Eq.~\eqref{mrfm::eq:H:3}.  In this section, we now take
into account the influence of the environment starting from the
Hamiltonian Eq.~\eqref{mrfm::eq:H:total}.  The reduced dynamics is
obtained analytically with the Feynman-Vernon influence
functional\cite{feynmanvernon,grabert} for arbitrary coupling strength
\(\alpha\) to the bath and for arbitrary temperature \(T\). The
advantage of this method as compared Ref.~\onlinecite{Berman03b} 
is that no master equation is used and that
there is no restriction on the number of basis functions used to
numerically integrate the problem.

The reduced dynamics of the cantilever obtained with the influence
functional is given by:
\begin{eqnarray}
&&\rho_{s s'}^{(C)}(z_{f},z'_{f},t)=
\\\nonumber
&&=\int dz_{i}dz'_{i}J_{s s'}(z_{f},z'_{f},t;z_{i},z'_{i},t_{0})
\rho_{ss'}^{(C)}(z_{i},z'_{i},t_{0})\,,
\end{eqnarray}
where the influence functional is
\begin{equation}
J_{s s'}(z_{f},z'_{f},t;z_{i},z'_{i},t_{0})=
\int \varD z \varD z'\exp(i S_{s s'}[z,z'])
\,,
\end{equation}
\(s,s'=\pm\), and the action is defined by:
\begin{eqnarray}
&&S_{s s'}[z,z']=S_{s}^{0}[z]-S_{s'}^{0}[z']
\\\nonumber
&&-\frac{\alpha}{2}\int_{t_{0}}^{t}d\tau[z(\tau)-
z'(\tau)][\dot{z}(\tau)+\dot{z'}(\tau)]
\\\nonumber
&&+\frac{i}{2}\int_{t_{0}}^{t}d\tau\int_{t_{0}}^{t}d\tau'
[z(\tau)-z'(\tau)]K(\tau-\tau')[z(\tau')-z'(\tau')]
\,.
\end{eqnarray}
This form of the action is only valid for an Ohmic bath\cite{Caldeira83b}.
Furthermore, $K(\tau)$ is the real part of the bath correlation function
\begin{equation}
K(\tau)\equiv\re\langle \hat{X}(\tau)\hat{X}(0)\rangle\,,
\end{equation}
where $\hat{X}(t)=\sum_k c_k \hat{x}_k(t)$.
Finally, 
\begin{equation}
S_{s}^{0}[z]=\int_{t_{0}}^{t}d\tau[\frac{1}{2}\dot{z}^2
(\tau)-\frac{1}{2}z^2(\tau)+\eta s
f(\tau)z(\tau)+\frac{1}{2}\epsilon(\tau)s]\,
\end{equation}
is the bare action without oscillator bath.

The action can be simplified further by introducing relative
coordinates defined by \(R=(z+z')/2\) and \(r=z-z'\). The action is
then found to be
\begin{eqnarray}\nonumber
&&S_{s s'}[R,r]=S_{s
s'}^{0}[R,r]-\alpha\int_{t_{0}}^{t}d\tau\dot{R}(\tau)r(\tau)
\\
&&+\frac{i}{2}\int_{t_{0}}^{t}d\tau\int_{t_{0}}^{t}d\tau' 
r(\tau)K(\tau-\tau') r(\tau')\,,
\end{eqnarray}
with
\begin{eqnarray}\nonumber
&&S_{s s'}^{0}[R,r] = \\
\nonumber
&&=\int_{t_{0}}^{t}d\tau\Big\{\dot{R}(\tau)\dot{r}(\tau)-
R(\tau)r(\tau)+\eta f(\tau)R(\tau)(s-s')
\\
&&+\frac{1}{2}\eta
f(\tau)r(\tau)(s+s')+\frac{1}{2}\epsilon(\tau)(s-s')\Big\}\,.
\end{eqnarray}

In the next step, the action is expanded around the classical path.
The classical equations of motion can be found by minimizing this
action and read
\begin{equation}
\ddot{R}(\tau)+\alpha\dot{R}(\tau)+R(\tau)=F_{R}(\tau)\,,
\label{classR}
\end{equation}
\begin{equation}
\ddot{r}(\tau)-\alpha\dot{r}(\tau)+r(\tau)=F_{r}(\tau)\,,
\label{classr}
\end{equation}
\begin{equation}
F_{R}(\tau)=\frac{1}{2}\eta f(\tau)(s+s')+
i\int_{t_{0}}^{t}d\tau'K(\tau-\tau')r(\tau')\,,
\end{equation}
\begin{equation}
F_{r}(\tau)=\eta f(\tau)(s-s')\,,
\end{equation}
with classical solutions \(R_{cl}(\tau)\), \(r_{cl}(\tau)\),
respectively. Note that the solutions are complex \cite{Marquardt02},
and the dependence on \(s,s'\) of all these
quantities has been suppressed. The classical solutions, which are
given in Appendix~\ref{mrfm::sec:path}, are linear in the boundary
values \(R_{f},r_{f},R_{i}\) and \(r_{i}\). Therefore, 
\(S_{s s'}[R_{cl},r_{cl}]\) is a bilinear form in these variables.
We obtain
\begin{equation}
J_{s s'}(R_{f},r_{f},t;R_{i},r_{i},t_{0})=\frac{1}{\mathcal{N}(t)}\exp\Big(i
S_{s s'}[R_{cl},r_{cl}]\Big)\,,
\end{equation}
where all the contributions from the fluctuations around the classical
path are contained in the time-dependent, but
spin-independent normalization constant \(\mathcal{N}(t)\), which can
be obtained from the normalization condition
\begin{equation}
\sum_{s=\pm}\int_{-\infty}^{\infty} dR_{f}\rho_{s s}(R_{f},r_{f}=0,t)=1\,.
\end{equation}
The Gaussian form of the expressions leads to a final reduced density
matrix of Gaussian form if the initial density matrix is Gaussian,
which is true for a coherent state. Therefore we deal with Gaussian
wave packets also in the dissipative case. The explicit formulas are
discussed in detail in Appendix~\ref{mrfm::sec:path}, where the
solution for the reduced dynamics is obtained starting from a Gaussian
wave packet at time \(t_{0}\). We then take the limit
\(t_{0}\rightarrow -\infty\) such that the information about the initial
state is lost at time \(t=0\).

We will now give analytical expressions of the density matrix for
the diagonal and off-diagonal elements with respect to the spin degree
of freedom.  Let us first discuss the result for $s=s'$:
\begin{eqnarray}\label{mrfm::eq:diag}
\nonumber
\rho_{s s}^{(C)}(R,r,t)=\frac{1}{\sqrt{2\pi}\sigma_{R}}
\exp\{-\frac{1}{2 \sigma_{R}^2}[R-x_{s}(t)]^2\\
-\frac{1}{2\sigma_{r}^2}r^2+ir\dot{x}_{s}(t) \}\,,
\end{eqnarray}
where the final coordinates have been replaced by \(R\equiv R_{f}\)
and \(r\equiv r_{f}\).
The widths of the Gaussian peaks are
independent of the spin. 
The width in the \(R\)-direction is given by
\begin{equation}\label{mrfm::eq:delR}
\sigma_{R}^2=\int_{0}^{\infty}d\omega
J_{\eff}(\omega)\coth\Big(\frac{\omega}{2 T}\Big)\,.
\end{equation}
 $\sigma_{R}$ increases with temperature.  This is because
the cantilever position suffers more thermal fluctuations.
The width in \(r\)-direction is found to be
\begin{equation}\label{mrfm::eq:delr}
\frac{1}{\sigma_{r}^2}=\int_{0}^{\omega_C}d\omega \omega^2
J_{\eff}(\omega)\coth\Big(\frac{\omega}{2 T}\Big)\,.
\end{equation}
Note that as is well-known the momentum width diverges with the
cut-off frequency \(\omega_C\) which was defined after
Eq.~\eqref{mrfm::eq:J}. That is why we kept the dependence on the
cut-off in this integral. The spin dynamics and the probability
distribution of the cantilever will not depend on the cut-off. In
contrast to $\sigma_{R}$, $\sigma_{r}$ decreases with temperature;
this is natural since the cantilever gets closer to a classical
oscillator as temperature goes up.  The temperature behavior of these
two integrals can be read off in the limit of small \(\alpha\ll
1\), viz.,
\begin{equation}
\sigma_{R}^2\approx\frac{1}{\sigma_{r}^2}\approx\frac{1}{2}
\coth\Big(\frac{1}{2 T}\Big)\,.
\end{equation}

The Gaussian wave packets are moving according to
\begin{equation}
x_{s}(t)=\eta s\int_{0}^{t}dt'e^{-\frac{\alpha}{2}(t-t')}
\frac{\sin\Big(\omega_{R}(t-t')\Big)}{\omega_{R}}f(t')\,,
\end{equation}
which depends on the spin \(s\).  The oscillator frequency
\(\omega_{R}=\sqrt{1-(\alpha/2)^2}\) is renormalized due to the
coupling to the bath.  Furthermore, \(x_{s}(t)\) is the solution of
the coordinate of a classical dissipative driven harmonic oscillator
with a spin-dependent driving force \(\eta s f(t)\) starting from the
initial conditions \(x_{s}(0)=0\) and \(\dot{x}_{s}(0)=0\). So the result
becomes very clear, because the classical solution is well-known to be
an oscillating function, which goes trough a transient regime and for
$t\gg 1/\alpha$ the amplitude of the oscillation saturates at a finite
value. The oscillation is periodic (but not necessarily sinusoidal) in
time with unit period ($T_0=2\pi/\omega_0$).  Consequently, for $t\gg
1/\alpha$ the density matrix will show a generic steady-state behavior
independent of the details of the initial preparation of the system.

\begin{figure}
\centering
\includegraphics[width=.8\linewidth]{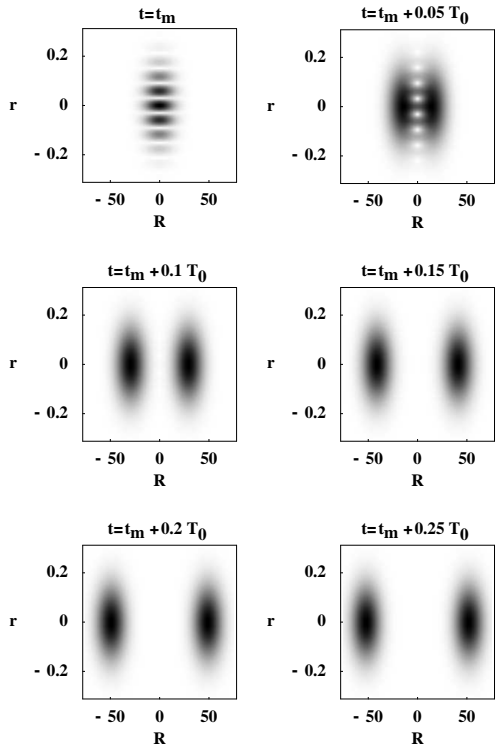}
\caption{\label{mrfm::fig:3} \(|\rho^{(C)}(R,r,t)|\) 
for a time series in the steady-state regime starting
at time \(t_{m}\) at which the two peaks are not separated,
e.g., \(t_{m}=988\)
The units have been chosen such that both the natural frequency
$\omega_0$ of the
cantilever and its harmonic oscillator length are equal to 1.
$T_0=2\pi/\omega_0$, \(\alpha=0.006\), \(T=100\); 
the other parameters are as in the caption of Fig.~\ref{mrfm::fig:2}. 
The interference fringes are due to the driving.
}
\end{figure}

\begin{figure}
\centering
\includegraphics[width=.8\linewidth]{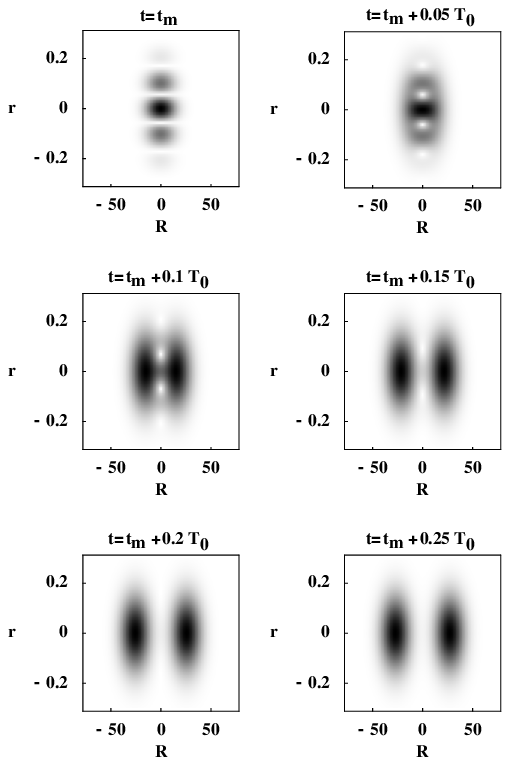}
\caption{\label{mrfm::fig:4} \(|\rho^{(C)}(R,r,t)|\) 
for a time series in the steady-state regime for
  \(\alpha=0.012\) and \(T=100\).}
\end{figure}

\begin{figure}
\centering
\includegraphics[width=.8\linewidth]{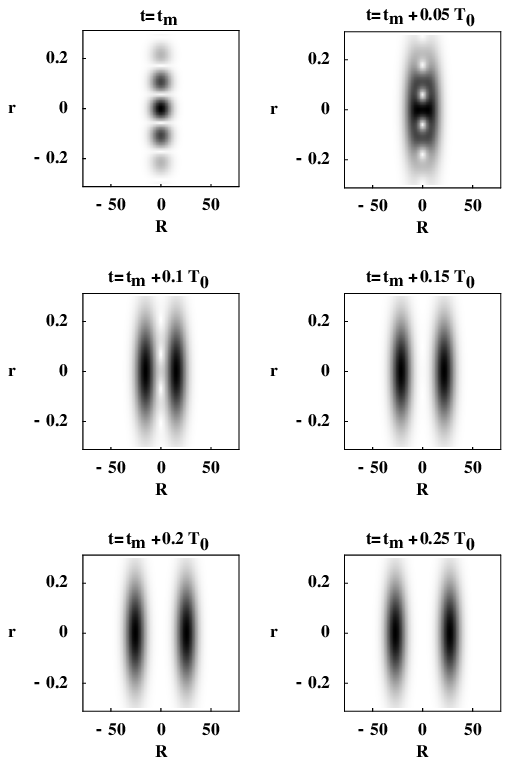}
\caption{\label{mrfm::fig:5} \(|\rho^{(C)}(R,r,t)|\) 
for a time series in the steady-state regime for
  \(\alpha=0.012\) and \(T=50\).}
\end{figure}

\begin{figure}
\centering
\includegraphics[width=.8\linewidth]{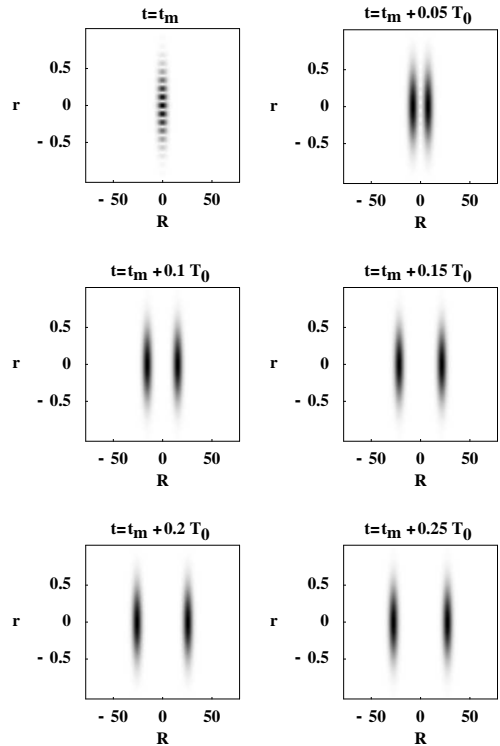}
\caption{\label{mrfm::fig:6} \(|\rho^{(C)}(R,r,t)|\) 
for a time series in the steady-state regime for
  \(\alpha=0.012\) and \(T=10\).}
\end{figure}

The density matrix $\rho_{ss}^{(C)}(R,r,t)$ behaves quite
differently with respect to the coordinates $R$ and $r$. As a function
of $R$, $\rho_{ss}^{(C)}(R,r,t)$ is a Gaussian distribution with
standard deviation \(\sigma_{R}\) and average \(\avg{R(t)}=x_{s}(t)\).
On the other hand, $\dot{x}_{s}(t)$ is the velocity of a classical
oscillator [see above], it shows oscillatory behavior in $t$ and $r$
superimposed on the Gaussian envelope with width \(\sigma_{r}\); see
Figs.~\ref{mrfm::fig:3}--\ref{mrfm::fig:6}.  Thus, the off-diagonal
elements $\rho_{ss}^{(C)}(z,z',t)$ ($z\neq z'$) exhibit an oscillating
behavior in $t$.  However, this should not be confused with a coherent
oscillation, which is not expected in this long-time limit.  The
oscillation is a consequence of the external driving $f(t)$ [i.e.,
frequency modulation $\dot\phi(t)$].  The diagonal elements
(both in $s$ and $z$) $\rho_{ss}^{(C)}(z,z,t)$ do not show such an
oscillation.

The behavior of \(x_{s}(t)\) can be illustrated by approximating
\(f(t)\) by its primary oscillation amplitude:
\begin{equation}
f(t)\approx f_0 \sin(t) + (\text{higher harmonics})\,,
\end{equation}
where
\begin{equation}
f_0 = \frac{4}{\pi}\left(\frac{\eps_\bot}{\phi_0}\right)
\left[E(-\phi_0^2/\eps_\bot^2) - K(-\phi_0^2/\eps_\bot^2)\right]\,.
\end{equation}
Here, $K(x)$ and $E(x)$ are the complete elliptic integrals of the first
and second kind~\cite{Abramowitz72a}.
One obtains 
\begin{eqnarray}\nonumber
\label{mrfm::eq:R(t)}
x_{s}(t)&\approx& \eta s
f_0\big(-\frac{\cos(t)}{\alpha}+e^{-\frac{\alpha}{2}t}
[\frac{\cos(\omega_{R}t)}{\alpha}+\frac{\sin(\omega_{R}t)}
{2\omega_{R}}]\big)
\\ &&+ (\text{higher harmonics})\,.
\end{eqnarray}
This solution shows the main features of the spin-dependent separation
\(x_{s}(t)\), namely the transient behavior and the steady-state
oscillations: \(x_{s}(t)\approx -\eta s f_0\cos(t)/\alpha\).  It is
interesting to notice that the average cantilever motions are exactly
in opposite phases (shift by $\pi$) for spin up ($s=+1$) and down
($s=-1$).  This was also concluded from the numerical simulation
presented Refs.~\onlinecite{Berman03a,Berman03b}.  Thus, the MRFM can
be used as a quantum measurement device, i.e., to detect
the state of the spin; see below.  Therefore, if we
start initially with the two spin components populated,
$\rho_{++}^{(S)}(0),\rho_{--}^{(S)}(0) > 0$, then
\begin{math}
\rho^{(C)}(R,r,t) = \rho_{++}^{(S)}(0)\rho_{++}^{(C)}(R,r,t) +
\rho_{--}^{(S)}(0)\rho_{--}^{(C)}(R,r,t)
\end{math}
will show two peaks moving in opposite directions as time goes on; see
discussions above and Figs.~\ref{mrfm::fig:3}--\ref{mrfm::fig:6}.  It
should be stressed that to separate the two peaks with sufficient
resolution, the widths of the peaks, Eq.~\eqref{mrfm::eq:delR}, should
not be larger than the maximum separation, $\eta f_0/\alpha$; see
Eq.~\eqref{mrfm::eq:R(t)}.  Clearly, this criterion restricts the
maximum operation temperature of the device.
Figures~\ref{mrfm::fig:3}--\ref{mrfm::fig:6} show the typical
behavior of the density matrix $\rho^{(C)}(R,r,t)$ of the cantilever
for $\rho_{ss'}^{(S)}(0)=1/2$ for $s,s'=\pm$ as initial state.
As the coupling to the
environment $\alpha$ increases, the distance between the peaks shrinks
and they are harder to distinguish; see Figs.~\ref{mrfm::fig:3} and
\ref{mrfm::fig:4}.  A similar behavior is observed as the temperature
increases with $\alpha$ fixed; see Figs.~\ref{mrfm::fig:5} and
\ref{mrfm::fig:6}.

Now we turn to the off-diagonal elements $s=-s'$:
\begin{eqnarray}\label{mrfm::eq:offdiag}\nonumber
&&\rho_{s,-s}^{(C)}(R,r,t)
=\frac{1}{\sqrt{2\pi}\sigma_{R}}\exp\{
-\frac{1}{2 \sigma_{R}^2}[R-i\vartheta_{s}(t)]^2\\
&&-\frac{1}{2 \sigma_{r}^2}r^2+r\zeta_{s}(t)
-\Gamma(t)+ i\int_{0}^{t}dt'\epsilon(t')\}\,,
\end{eqnarray}
where
\begin{eqnarray}
&&\vartheta_{s}(t)=2\eta s\int_{0}^{\infty}d\omega
J_{\eff}(\omega)\coth\Big(\frac{\omega}{2T}\Big)
\\\nonumber
&&\times\int_{0}^{t}dt' f(t')\cos\big(\omega(t-t')\big)\,,
\end{eqnarray}
and
\begin{eqnarray}
&&\zeta_{s}(t)=2\eta s\int_{0}^{\infty}d\omega
\omega J_{\eff}(\omega)\coth\Big(\frac{\omega}{2T}\Big)
\\\nonumber
&&\times\int_{0}^{t}dt' f(t')\sin\big(\omega(t-t')\big)\;.
\end{eqnarray}

In $r$-direction, $\rho_{s,-s}^{(C)}(R,r,t)$ has a Gaussian shape centered at
\(\zeta_{s}(t)/\sigma_{r}^{2}\) with width \(\sigma_{r}\).
In $R$-direction, it is an
oscillatory function imposed on a Gaussian envelope with width
\(\sigma_{R}\). Overall, the
function $\rho_{s,-s}^{(C)}(R,r,t)$ decays with $t$ in the same way as shown in
Fig.~\ref{mrfm::fig:2}, i.e.,  $\rho_{ss'}^{(C)}(R,r,t)$ for
$s\neq s'$ can be observed only in the transient regime.  The decay is
described by the function \(\Gamma (t)\); see
Eq.~\eqref{mrfm::eq:Gamma(t):2}. Note that a trace over the cantilever
dynamics leads us back to the results obtained in a much simpler way
in Section~\ref{mrfm::sec:spin}.

\section{MRFM as a Quantum Measurement Device}%
\label{mrfm::sec:qmeasure}%

One of the conclusions of the analysis presented here is that the
cantilever oscillates with the same amplitude for both initial spin
states (up and down).  Probing the amplitude of the cantilever
vibration can only tell the 
absolute value of the spin in the direction of $\bfB_\eff(0)$,
but not its sign. However, the oscillations for the initial
spin up and down states are completely out of phase (phase difference
of $\pi$); see Section \ref{mrfm::sec:cantilever}.  This fact was also
noticed by Berman \textit{et al.}~\cite{Berman03a} in their numerical
simulations. Hence, there is the possibility to use the MRFM as a quantum
measurement device, i.e., to detect the direction of the spin
with the MRFM by probing the (discrete) relative phases of the
cantilever oscillations.  In the quantum theory of measurement, this
falls into the category of the indirect quantum measurement
scheme~\cite{Breuer02a}.  In such a scheme, the quantum object supposed
to be measured is coupled to another quantum system, the so-called
quantum probe.  The classical measurement device then detects the
quantum probe instead of probing directly the quantum object.
In our case, the quantum object is the spin, the quantum
probe corresponds to the cantilever, and the classical measurement
device can be, e.g., the fiber-optical interferometer.  A conceivable
scheme to measure the relative phases of the cantilever oscillations
is to use a reference spin which is prepared in a definite known
state, for example, by applying a strong magnetic field in a desired
direction.  The two signals from the reference spin and the spin in an
unknown state are superposed to determine the relative phase of the unknown
spin.

\section{Conclusions}
\label{mrfm::sec:conclusion}

We have studied the CAI-based MRFM as a high-resolution tool to detect
single spins.  The quantum dynamics of the spin-plus-cantilever system
was analyzed in terms of the reduced density matrices,
$\hat\rho^{(S)}(t)$ (for the spin) and $\hat\rho^{(C)}(t)$ (for the
cantilever), in the presence of coupling to the environment.  Using an
effective bath model, we were able to determine the dynamics of the
spin during the measurement process.  Our results remain valid at all
temperatures as long as the adiabatic approximation is satisfied.  We
have evaluated the influence functional for the combined system of spin
and cantilever to obtain the quantum dissipative dynamics of the
cantilever.  These results are valid for all temperatures and coupling
strengths.  Finally, we have proposed that the MRFM can be used 
as a quantum measurement device, i.e., not
only to detect the absolute value of the spin but also to detect its 
direction.

The dissipative dynamics of an open quantum system is sensitive to the
low-frequency behavior of the spectral density of the environment.
While the Ohmic model Eq.~\eqref{mrfm::eq:J} is a plausible model, it
will be worthwhile to identify the sources of the environmental
fluctuations and construct a physical model of the environment
starting from a more microscopic theory of the cantilever.

\begin{acknowledgments}
We would like to thank V.\ Cerletti, F.\ Meier, and particularly F.\
Marquardt for helpful discussions. Our work was supported by the
SKORE-A program, the Swiss NSF, and the NCCR Nanoscience.
H.G.\ and M.-S.C.\ acknowledge the support by KIAS, 
where part of the work was done.  M.-S.C.\ has been supported by
a Korea Research Foundation Grant (KRF-2002-070-C00029).  
\end{acknowledgments}

%%%%
\appendix
\section{Estimation of the spin-flip rate}
\label{mrfm::sec:Landau-Zener}

The cyclic adiabatic inversion scheme implies two basic assumptions: (i)
The variation of the external driving $\dot\phi(t)$ is slow enough to
allow for an adiabatic approximation~\cite{Messiah61b}, i.e.,
\begin{math}
\left|\ddot\phi(t)\right| \ll \eps_\bot^2
\end{math}.
(ii) The time scales of the spin dynamics and the cantilever dynamics
are well separated ($\eps_\bot\gg1$) such that the Born-Oppenheimer
approximation is justified.  Yet, the finite rates of change in the
external driving and the cantilever position will induce spin flips.  The
spin-flip rate can be estimated by the Landau-Zener transition
(adiabatic transition)
rate~\cite{Avron87a,Grifoni98a,Landau32a,Zener32a,Stuckelberg32a}.  For
this purpose, we rewrite Eqs.~\eqref{mrfm::eq:H:1} and
\eqref{mrfm::eq:H:total} in the form
\begin{equation}
\varH_\mathrm{LZ}(t)
= -\frac{1}{2}F(t)\hat{\sigma}_{z}
- \frac{1}{2}\eps_\bot\hat{\sigma}_{x}\,,
\end{equation}
where
\begin{math}
F(t)
\equiv \dot{\phi}(t)
+ 2\eta\left\langle\hat{z}(t)\right\rangle
\end{math}.
The back-action of the cantilever has been accounted for by its
time-dependent average position, and the contribution from it will be
estimated below in a self-consistent way based on the results in
Section~\ref{mrfm::sec:cantilever}. The probability that the spin flips
against the effective magnetic field $\bfB_\eff(t)$ during one period
(i.e., $2\pi/\omega_0$) is then given by
\begin{equation}
\label{mrfm::eq:PLZ}
P_\mathrm{LZ}
\simeq \exp(-\frac{\pi\eps_\bot^2}{\nu}) \,,
\end{equation}
where we have taken
\begin{math}
\nu \equiv \max\left|\dot F(t)\right|
\end{math}
to estimate the worst case.

It follows from Eqs.~\eqref{mrfm::eq:dotphi} and \eqref{mrfm::eq:R(t)}
that
\begin{equation}
\nu \leq \max\left|\ddot\phi(t)\right|
+ 2\eta\max\left|\frac{d}{dt}\avg{\hat{z}(t)}\right|
= \phi_0 + 2\frac{\eta^2}{\alpha}f_0 \,.
\end{equation}
Therefore, assuming typical values for the parameters,
$\phi_0\sim 1000$, $\eps_\bot\sim 400$, $\eta\sim 1$, and $\alpha\sim
0.001$, we have $f_0\sim 1$ and
\begin{equation}
P_\mathrm{LZ}
< \exp\left(-\pi\frac{\eps_\bot^2}{\phi_0+2\eta^2f_0/\alpha}\right)
\sim 10^{-70} \,.
\end{equation}
Note that the back-action of the cantilever is stronger for larger
$Q$-factors of the cantilever ($Q\simeq 1/\alpha$) since the maximum
velocity of the cantilever increases with the $Q$-factor.

\section{Path-integral formulas}
\label{mrfm::sec:path}
\begin{widetext}
In this appendix we will fill in some of the details left out in 
Section~\ref{mrfm::sec:cantilever}. It is convenient to define 
\(\gamma\equiv\alpha/2\) as the friction constant.
The classical solutions to Eqs.~(\ref{classR}), (\ref{classr})
are given by
\begin{equation}
r_{cl}(\tau)=\frac{1}{\sin\big(\omega_{R}(t-t_{0})\big)}\Big\{r_{i}
\sin\big(\omega_{R}(t-\tau)\big)e^{\gamma(\tau-t_{0})}+
[r_{f}-r_p(t)]
\sin\big(\omega_{R}(\tau-t_{0})\big)e^{\gamma(\tau-t)}\Big\}
+r_{p}(\tau)\,,
\end{equation}

\begin{equation}
R_{cl}(\tau)=\frac{1}{\sin\big(\omega_{R}(t-t_{0})\big)}\Big\{R_{i}
\sin\big(\omega_{R}(t-\tau)\big)e^{-\gamma(\tau-t_{0})}
+[R_{f}-R_{p}(t)]\sin\big(\omega_{R}(\tau-t_{0})\big)e^{-\gamma(\tau-t)}\Big\}
+R_{p}(\tau)\,,
\end{equation}
where
\begin{equation}
r_{p}(\tau)=\int_{t_{0}}^{\tau}d\tau'G_{r}(\tau-\tau')F_{r}(\tau')\,,
\end{equation}
\begin{equation}
R_{p}(\tau)=\int_{t_{0}}^{\tau}d\tau'G_{R}(\tau-\tau')F_{R}(\tau')\,,
\end{equation}
and the Green's functions are defined by
\begin{equation}
G_{R}(\tau)=\Theta(\tau)e^{-\gamma\tau}\frac{\sin(\omega_{R}\tau)}
{\omega_{R}}\,,
\end{equation}
\begin{equation}
G_{r}(\tau)=\Theta(\tau)e^{\gamma\tau}\frac{\sin(\omega_{R}\tau)}
{\omega_{R}}\,.
\end{equation}
The influence functional for  \(s'=s\) is found to be
\begin{eqnarray}
&&J_{s
s}(R_{f},r_{f},t;R_{i},r_{i},t_{0})=
\\\nonumber
&&\frac{|N(t)|}{2\pi}\exp\Big(i[K_{f}(t)R_{f}r_{f}+K_{i}(t)R_{i}r_{i}-
L(t)R_{i}r_{f}-N(t)R_{f}r_{i}+a_{i}(t)r_{i}+a_{f}(t)r_{f}]-
A(t)r_{f}^2-B(t)r_{f}r_{i}-C(t)r_{i}^2\Big)\,,
\end{eqnarray}
where the functions appearing in the influence functional are all real and
defined by
\begin{equation}
K_{f}(t)=\omega_{R}\cot\big(\omega_{R}(t-t_{0})\big)-\gamma\,,
\end{equation}
\begin{equation}
K_{i}(t)=\omega_{R}\cot\big(\omega_{R}(t-t_{0})\big)+\gamma\,,
\end{equation}
\begin{equation}
L(t)=\frac{\omega_{R}e^{-\gamma(t-t_{0})}}{\sin\big(\omega_{R}(t-t_{0})\big)}
\,,
\end{equation}
\begin{equation}
N(t)=\frac{\omega_{R}e^{\gamma(t-t_{0})}}{\sin\big(\omega_{R}(t-t_{0})\big)}
\,,
\end{equation}
\begin{equation}
A(t)=\frac{1}{2}\frac{e^{-2\gamma t}}{\sin^2\big(\omega_{R}(t-t_{0})\big)}
\int_{t_{0}}^{t}d\tau\int_{t_{0}}^{t}d\tau'\sin\big(\omega_{R}(\tau-t_{0})\big)
K(\tau-\tau')\sin\big(\omega_{R}(\tau'-t_{0})\big)e^{\gamma(\tau+\tau')}\,,
\end{equation}
\begin{equation}
B(t)=\frac{e^{-\gamma (t+t_{0})}}{\sin^2\big(\omega_{R}(t-t_{0})\big)}
\int_{t_{0}}^{t}d\tau\int_{t_{0}}^{t}d\tau'\sin\big(\omega_{R}(t-\tau)\big)
K(\tau-\tau')\sin\big(\omega_{R}(\tau'-t_{0})\big)e^{\gamma(\tau+\tau')}\,,
\end{equation}
\begin{equation}
C(t)=\frac{1}{2}\frac{e^{-2\gamma t_{0}}}{\sin^2\big(\omega_{R}(t-t_{0})\big)}
\int_{t_{0}}^{t}d\tau\int_{t_{0}}^{t}d\tau'\sin\big(\omega_{R}(t-\tau)\big)
K(\tau-\tau')\sin\big(\omega_{R}(t-\tau')\big)e^{\gamma(\tau+\tau')}\,,
\end{equation}
\begin{equation}
a_{f}(t)=\dot{x}(t)-K_{f}(t)x(t)\,,
\end{equation}
\begin{equation}
a_{i}(t)=N(t)x(t)\,,
\end{equation}
\begin{equation}
x(\tau)=\eta s\int_{t_{0}}^{\tau}d\tau'G_{R}(\tau-\tau')f(\tau')\,,
\end{equation}
\begin{equation}
\dot{x}(\tau)=\eta s\int_{t_{0}}^{\tau}d\tau'\partial_{\tau}G_{R}(\tau-\tau')
f(\tau')\,.
\end{equation}
In all of these expressions, the dependence on \(t_{0}\) has been
suppressed.

Let us now discuss the solution for the density matrix.  At time
\(t=t_{0}\) we start in a product state between cantilever and
bath. The cantilever density matrix is assumed to be a Gaussian
wave packet with width \(\sigma\) at \(t=t_{0}\),
\begin{equation}
\rho_{ss'}^{(C)}(z,z',t_{0})=
\frac{1}{\sqrt{2\pi}\sigma}\exp\Big(-\frac{1}{4\sigma^2}(z^2+z'^2)\Big)
\,.
\end{equation}
One could start from a more general initial state, but we will later
take the limit \(t_{0}\rightarrow -\infty\), such that 
all the information on the initial state is lost completely at time \(t=0\).
The experiment starts at
time \(t=0\) by switching on the magnetic field. At this time the
cantilever has interacted with the bath for a very long time and is
in equilibrium with the bath, i.e., not any more in a product state.

The general solution for the diagonal elements of
$\rho_{s s'}^{(C)}$ starting from this initial condition is
\begin{eqnarray}\nonumber
&&\rho_{s s}^{(C)}(R_{f},r_{f},t)=
\frac{|N(t)|}{\sqrt{2\pi}}\frac{2\sigma}{\sqrt{D(t)}}
\\\nonumber
&&\times\exp\Bigg\{\Bigg[r_{f}^2\Bigg(-A(t)+[2 B^2(t)-8 A(t)C(t)-
\frac{L^2(t)}{2}]\sigma^2-
4\Big(A(t)K_{i}^2(t)+L(t)[B(t)K_{i}(t)+C(t)L(t)]\Big)\sigma^4\Bigg)
\\\nonumber
&&+i
r_{f}\Bigg(a_{f}(t)-4[a_{i}(t) B(t)-2a_{f}(t)C(t)]\sigma^2+4
K_{i}(t)[a_{f}(t) K_{i}(t)+a_{i}(t) L(t)]\sigma^4\Bigg)
\\\nonumber
&&+i R_{f}r_{f}\Bigg(K_{f}(t)+4[2
C(t)K_{f}(t)+B(t)N(t)]\sigma^2+4
K_{i}(t)[K_{f}(t)K_{i}(t)-L(t) N(t)]\sigma^4\Bigg)
\\
&&-2[a_{i}(t)-N(t)R_{f}]^2\sigma^2\Bigg]/D(t)\Bigg\}\,,
\end{eqnarray}
where
\begin{equation}
D(t)=1+8 C(t)\sigma^2+4K_{i}^2(t)\sigma^4\,.
\end{equation}

In the limit \(t_{0}\rightarrow -\infty\) we obtain the final result
presented in Eq.~(\ref{mrfm::eq:diag}).

The influence functional for \(s'=-s\) is found to be given by
\begin{eqnarray}
&&J_{s,-s}(R_{f},r_{f},t;R_{i},r_{i},t_{0})=
\\\nonumber
&&\frac{|N(t)|}{2\pi}\exp\Big(i[K_{f}(t)R_{f}r_{f}+
K_{i}(t)R_{i}r_{i}-L(t)R_{i}r_{f}-N(t)R_{f}r_{i}+
A_{f}(t)R_{f}+ A_{i}(t)R_{i}+\int_{t_{0}}^{t}d\tau\epsilon(\tau)]\Big)
\\\nonumber
&&\times\exp\Big(-A(t)r_{f}^2-B(t)r_{f}r_{i}-C(t)r_{i}^2+
b_{i}(t)r_{i}+b_{f}(t)r_{f}+b(t)\Big)\,,
\end{eqnarray}
where
\begin{equation}
A_{f}(t)=\dot{y}(t)-K_{i}(t)y(t)\,,
\end{equation}
\begin{equation}
A_{i}(t)=L(t)y(t)\,,
\end{equation}
\begin{equation}
b_{f}(t)=2
A(t)y(t)-\int_{t_{0}}^{t}d\tau\int_{t_{0}}^{t}d\tau'y(\tau')
K(\tau-\tau')\frac{\sin\big(\omega_{R}(\tau-t_{0})\big)
e^{-\gamma(t-\tau)}}{\sin\big(\omega_{R}(t-t_{0})\big)}\,,
\end{equation}
\begin{equation}
b_{i}(t)=B(t)y(t)-\int_{t_{0}}^{t}d\tau\int_{t_{0}}^{t}d\tau'y(\tau')
K(\tau-\tau')\frac{\sin\big(\omega_{R}(t-\tau)\big)
e^{\gamma(\tau-t_{0})}}{\sin\big(\omega_{R}(t-t_{0})\big)}\,,
\end{equation}
\begin{eqnarray}\nonumber
b(t)
=-A(t)y^2(t)+y(t)\int_{t_{0}}^{t}d\tau\int_{t_{0}}^{t}d\tau'y(\tau')
K(\tau-\tau')\frac{\sin\big(\omega_{R}(\tau-t_{0})\big)
e^{-\gamma(t-\tau)}}{\sin\big(\omega_{R}(t-t_{0})\big)}
\\
-\frac{1}{2}\int_{t_{0}}^{t}d\tau\int_{t_{0}}^{t}d\tau'y(\tau)
K(\tau-\tau')y(\tau')\,,
\end{eqnarray}

\begin{equation}
y(\tau)=2\eta s\int_{t_0}^{\tau}d\tau'G_{r}(\tau-\tau')f(\tau')\,,
\end{equation}
\begin{equation}
\dot{y}(\tau)=2\eta s
\int_{t_0}^{\tau}d\tau'\partial_{\tau}G_{r}(\tau-\tau')
f(\tau')\,.
\end{equation}

This leads to the following general expression for the off-diagonal
elements of $\rho_{s s'}^{(C)}$: 
\begin{eqnarray}\nonumber
&&\rho_{s,-s}^{(C)}(R_{f},r_{f},t)=\frac{|N(t)|}{\sqrt{2\pi}}\frac{2\sigma}
{\sqrt{D(t)}}
\\\nonumber
&&\times\exp\Bigg\{\Bigg[r_{f}^2\Bigg(-A(t)+[2 B^2(t)-8 A(t)C(t)-
\frac{L^2(t)}{2}]\sigma^2-4\Big(A(t)K_{i}^2(t)+L(t)[B(t)K_{i}(t)+
C(t)L(t)]\Big)\sigma^4\Bigg)
\\\nonumber
&&+r_{f}\Bigg(b_{f}(t)+[-4B(t)b_{i}(t)+8b_{f}(t)C(t)+A_{i}(t)L(t)]
\sigma^2
\\\nonumber
&&+4[A_{i}(t)B(t)K_{i}(t)+b_{f}(t)K_{i}^2(t)+2A_{i}(t)C(t)L(t)+b_{i}(t)
K_{i}(t)L(t)]\sigma^4\Bigg)
\\\nonumber
&&+i R_{f}r_{f}\Bigg(K_{f}(t)+4[2C(t)K_{f}(t)+B(t)N(t)]\sigma^2+4 
K_{i}(t)[K_{f}(t)K_{i}(t)-L(t)N(t)]\sigma^4\Bigg)
\\\nonumber
&&+i
R_{f}\Bigg(A_{f}(t)+8A_{f}(t)C(t)\sigma^2+4K_{i}(t)[A_{f}(t)K_{i}(t)+
A_{i}(t)N(t)]\sigma^4\Bigg)
\\\nonumber
&&+2[b_{i}(t)-iN(t)R_{f}]^2\sigma^2
\\\nonumber
&&-\frac{A_{i}^2(t)}{2}\sigma^2-4A_{i}(t)[A_{i}(t)C(t)+b_{i}(t)
K_{i}(t)]\sigma^4\Bigg]/D(t)
\\
&&+i\int_{t_{0}}^{t}d\tau\epsilon(\tau)+b(t)\Bigg\}\,.
\end{eqnarray}

The reduced dynamics of the spin alone is found by tracing out 
the cantilever coordinates.
The result is 
\begin{eqnarray}
\rho_{s,-s}^{(S)}(t)&=&\rho_{s,-s}^{(S)}(0)\exp
\big(-A_{f}^2(t)\frac{C(t)}{N^2(t)}+\frac{A_{f}(t)b_{i}(t)}
{N(t)}+b(t)-\frac{A_f^2(t)}{8\sigma^2N^2(t)}\\\nonumber
&&-\frac{\sigma^2}{2N^2(t)}[A_f(t)K_i(t)+A_i(t)N(t)]^2
+i\int_{t_{0}}^{t}d\tau\epsilon(\tau)
\big)
\\\nonumber
&&\equiv
\rho_{s,-s}^{(S)}(0)\exp\big(-\Gamma(t)+i\int_{t_{0}}^{t}
d\tau\epsilon(\tau)\big)\,.
\end{eqnarray}
The decay rate \(\Gamma(t)\); see Eq.~(\ref{mrfm::eq:Gamma(t):2}),
can be obtained in the limit \(t_{0}\rightarrow -\infty\) after a
straightforward but tedious calculation. In the same limit,
we get the result for the density matrix presented in
Eq.~(\ref{mrfm::eq:offdiag}).
\end{widetext}

%%%% References
%\bibliography{aliases,extra,mathey,cond-mat,stapkhy,physics,choims}%
%\bibliography{mrfm}

\end{document}